# Two Plasma Sources of Dayside Martian Magnetosphere: Pick-up Ions - Ionosphere Interaction.

O.Vaisberg, S.Shuvalov, A.Ramazan, V.Ermakov, I.Leonov

**Abstract**

Dayside magnetosphere of Mars is a thin layer with usually increased magnetic field and heated planetary ions between the magnetosheath and the ionosphere. The Mars Atmosphere and Volatile Evolution Mission (MAVEN) spacecraft provided, for the first time, high temporal and spatial resolution measurements enabling to study the structure of this region.

The region was given several names including the boundary layer, the plasma mantle, and magnetosphere. The outer boundary of this region was called magnetopause, ionopause, ion composition boundary and pile-up boundary (see Espley, 2018).

We analyze the properties and plasma populations in the layer between the magnetosheath flow and ionosphere on the day side of Mars. The region with the Solar-Zenith Angle (SZA) between $40^0$ and the subsolar point is considered. It was found that the region between magnetopause and ionosphere filled with the heated diluted ionospheric ions and the heavy pick-up ions is formed in magnetosheath. So, it is found that this region of heated ionospheric ion is collocated with another heavy ions component which has magnetosheath origin.

This magnetosheath component was eluded from identification due to two reasons. First one is that the second component looked like the high-energy tail of the heated ionospheric component on energy-time spectrograms. Second one is that STATIC energy-mass analyzer's energy range was shortened in the ionosphere diminishing observations of the higher energy ions. It appears that the magnetosheath ion component is important factor of formation of dayside magnetosphere and the source of energy in upper ionosphere and atmosphere.

**Introduction**

First investigations of the region with low energy plasma and increased magnetic field at dayside of Mars were performed on Mars-2 and Mars-3 and Mars-5 in $1970^{th}$ (Dolginov, 1976, Vaisberg, 1976, 1992), Phobos-2 (Szego, 1998) and MEX (Ramstad et al., 2017) provided comprehensive picture of nightside magnetosphere of Mars including the crustal fields in southern hemisphere of planet and the solar wind induced atmospheric losses.

Very detailed statistical analysis of the magnetosphere what authors name the Mars Composition Boundary Layer is presented in (Halekas et al., 2018)

MAVEN spacecraft with comprehensive high-time resolution instrument suite and favorable orbit gave possibility of detailed investigation of the Martian environment (Jakosky, 2015).

Analysis of Maven measurements at SZA ~ $70^o$ of Mars showed that the structure and properties of the dayside magnetosphere strongly depend on the location in MSE coordinate system. At "high magnetic latitudes" (small angle relative to direction of the motional electric field vector -1/c $V_{sw}$ x $B$ where $V_{sw}$ is the solar wind velocity vector) the energy of magnetospheric heavy ions is usually ~ 10-100 eV. Plume plasma often dominates in magnetosphere. Magnetic barrier is strong and spans through internal magnetosheath and magnetosphere. At "equatorial latitudes", where motional electric field is nearly horizontal, the main features are irregular structure of magnetic field in barrier and in magnetosphere, magnetospheric plasma is not steady and often consists of decreasing energy structures. At "low magnetic latitudes" where motional electric field is directed Mars-ward the magnetic barrier is weak and magnetospheric plasma has smooth structure and energy ~ 10-20 eV.

The model of magnetosphere formed by solar wind-planetary atmosphere interaction (usually called induced magnetosphere) (Vaisberg and Zelenyi, 1984, Zelenyi and Vaisberg 1986) indicates that the dayside magnetosphere forms from the narrow slab of incoming magnetic flux tubes entering the magnetosphere in subsolar region filled with pick-up ions and convection to the two flanks while continuing the loading of the heavy atmospheric ions

Analysis of the subsolar part of magnetosphere shows significant variability of the plasma and magnetic properties. There are several features that are different from what is observed at the magnetospheric flank..

**Observations**

For current analysis we selected 26 crossing of Martian magnetosheath-magnetosphere-ionosphere in the vicinity of subsolar region by MAVEN. Of these we selected 16 crossings during which orientation of STATIC energy-mass spectrometer was such that the direction to the Sun and direction of the spacecraft velocity were close to "equator" of its field-of-view. All passes of MAVEN considered were in the north hemisphere of Mars in order to minimize the influence of the crustal magnetic fields.

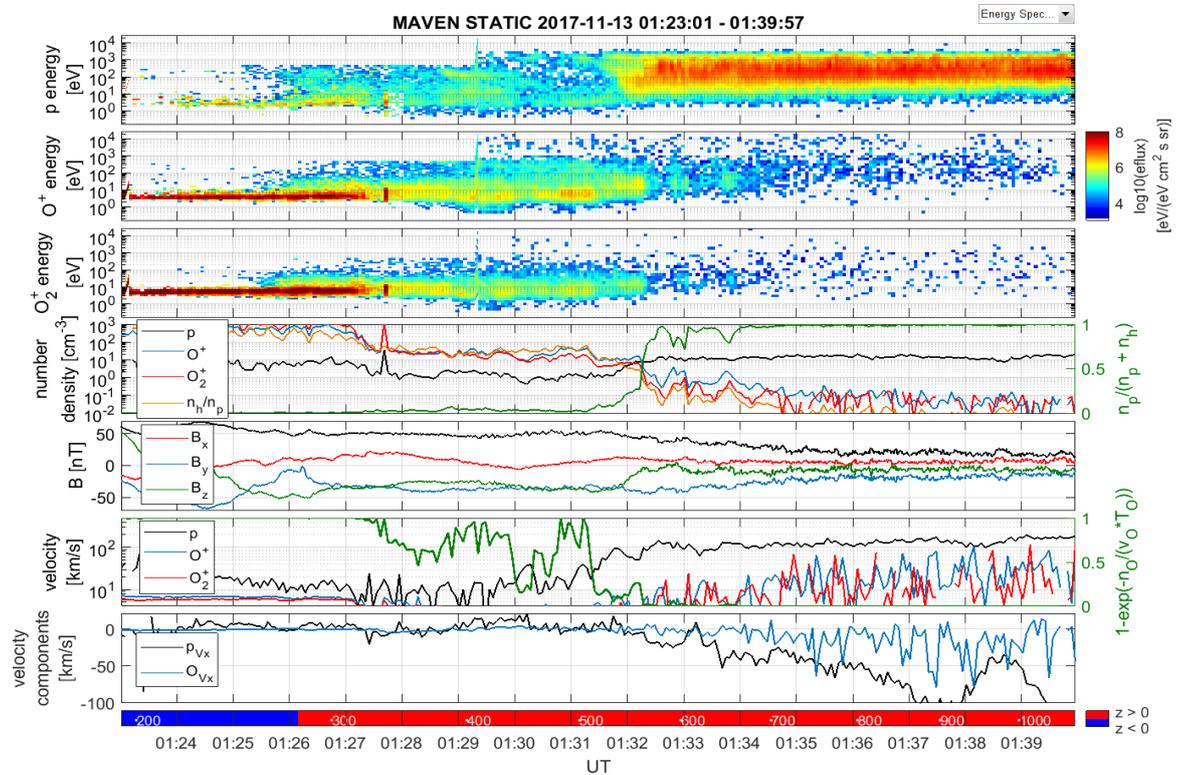

Figure 1. From top to bottom: the energy flux – spectrograms of protons, oxygen ions and ions $O^+_2$, number density of the same ions and ratio of protons to the sum of 2 oxygen ions

(green line), magnetic field components and magnitude (black line), the number density of 3 ions and the ration of O+ ions to the sum of their velocity and temperature.

Figure 1 shows crossing of ionosphere, magnetosphere and ionosheath on 13 November 2017 from 0123UT till 0140 UT. First, spacecraft crosses ionosphere (two thin red bands of $O^+$ and $O_2^+$ at low energies). At~ $01^m:25,30^s$ UT the ionospheric band widens and becomes green indicating decrease of the ionospheric ions number density. At~ 01:32 UT ionospheric number density drops and spacecraft enters ionosheath flow clearly seen as high-density energetic proton flux.

The blue lines at higher energies than main bands of ionospheric $O^+$ and $O_2^+$ start at about the time when main ionospheric component starts to widen, at ~ 01:25,30 UT. These ions are seen until magnetopause. After crossing the magnetopause the STATIC measures two wide bands of $O^+$ and $O_2^+$ that are pick-up ions moving to Mars with the solar wind. The density of these ions diminishes with distance from Mars expected due to diminishing exospheric neutral densities with the distance to Mars. The closest approach of MAVEN to Mars at altitude was 195 km at $01^h: 26^m$.at the solar-zenith angle of ~ $22^0$.

What is interesting is that the $O^+$ and $O_2^+$ blue bands in the upper part of ionosphere and in the pick-up region have quite similar energy and, possibly, widths. The explanation of these similarity will be proposed after consideration of other near-Mars space.

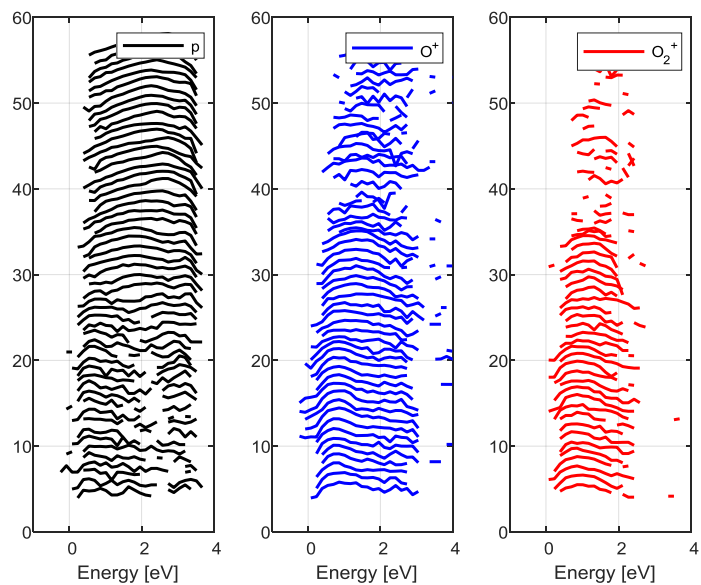

Figure 2. Spectra stack of ions in the time interval of 01:29 – 01: 34 UT.

Figure 2 shows profiles of the energy flux spectra in the time interval of 01:29 – 01: 34 UT. It is clearly seen two peaks of ions in stack of $O^+$ spectra indicating the higher energy band of ions is not the high-energy tail of ionospheric $O^+$ ions rather than the ions of another source.

Another pass of MAVEN near subsolar point at SZA angle ~ $6^0$ is shown in Figure 3. The measured properties of plasma are quite similar to previous case. Again one can see the ion component at higher energy range than ionospheric ions band that has similar energy range and similar umber density as the pick-up ion band in the solar wind flow. Note than the ion in the energy range <~500 eV the time interval < 08:02,40 UT were not measured in order to provide more detailed measurements of lower energies.

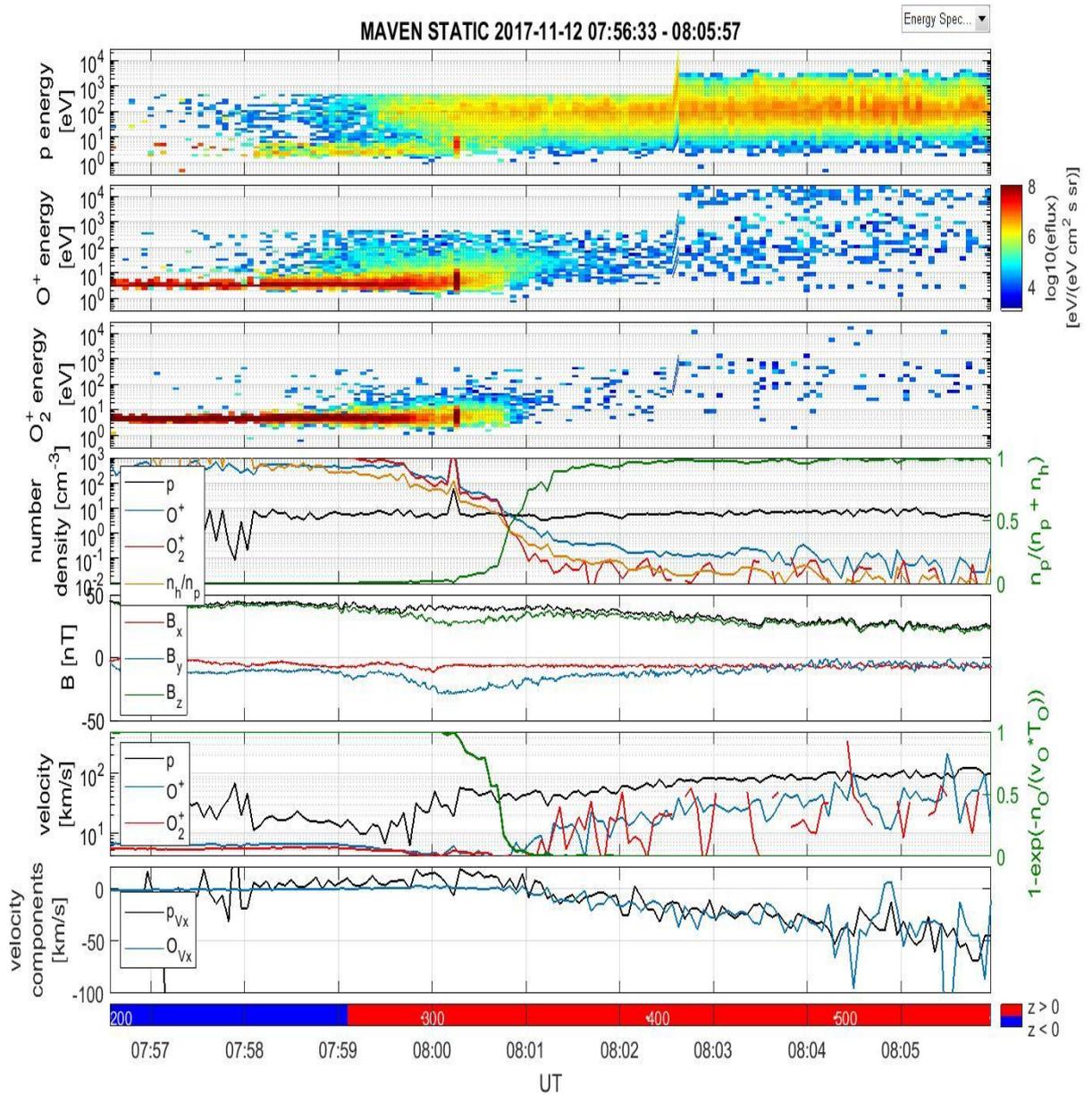

Figure 3. The pass of MAVEN in vicinity of subsolar point on November 12, 2017 at 07:57-08:06 UT.

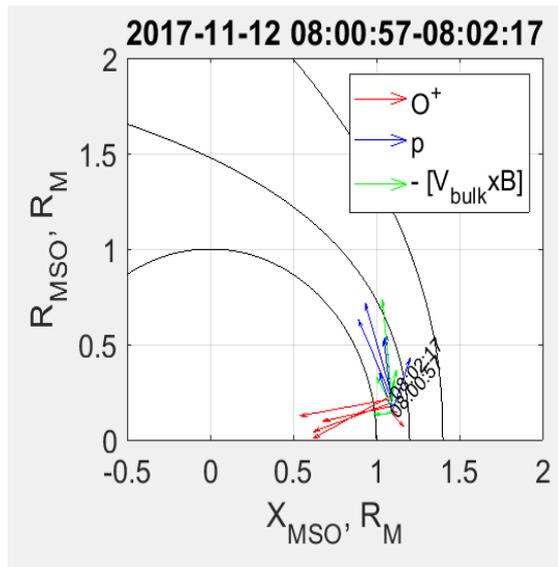

Figure 4. Electric field (green) vectors, O$^+$ (red) and proton (blue) velocities in cylindrical coordinate system.

Figure 4 shows short fraction of MAVEN orbit in cylindrical coordinate system on November 12$^{th}$ in the time interval 1$^m$20$^s$ starting at 08:00:57 the velocities of protons and O$^+$ ions and the motional electric field direction. One can see that protons are moving approximately in the direction of the magnetosheath flow while O$^+$ ions move downward to the planet. That indicates that what the velocity of heavy magnetosheath ions are moving within the magnetosphere.

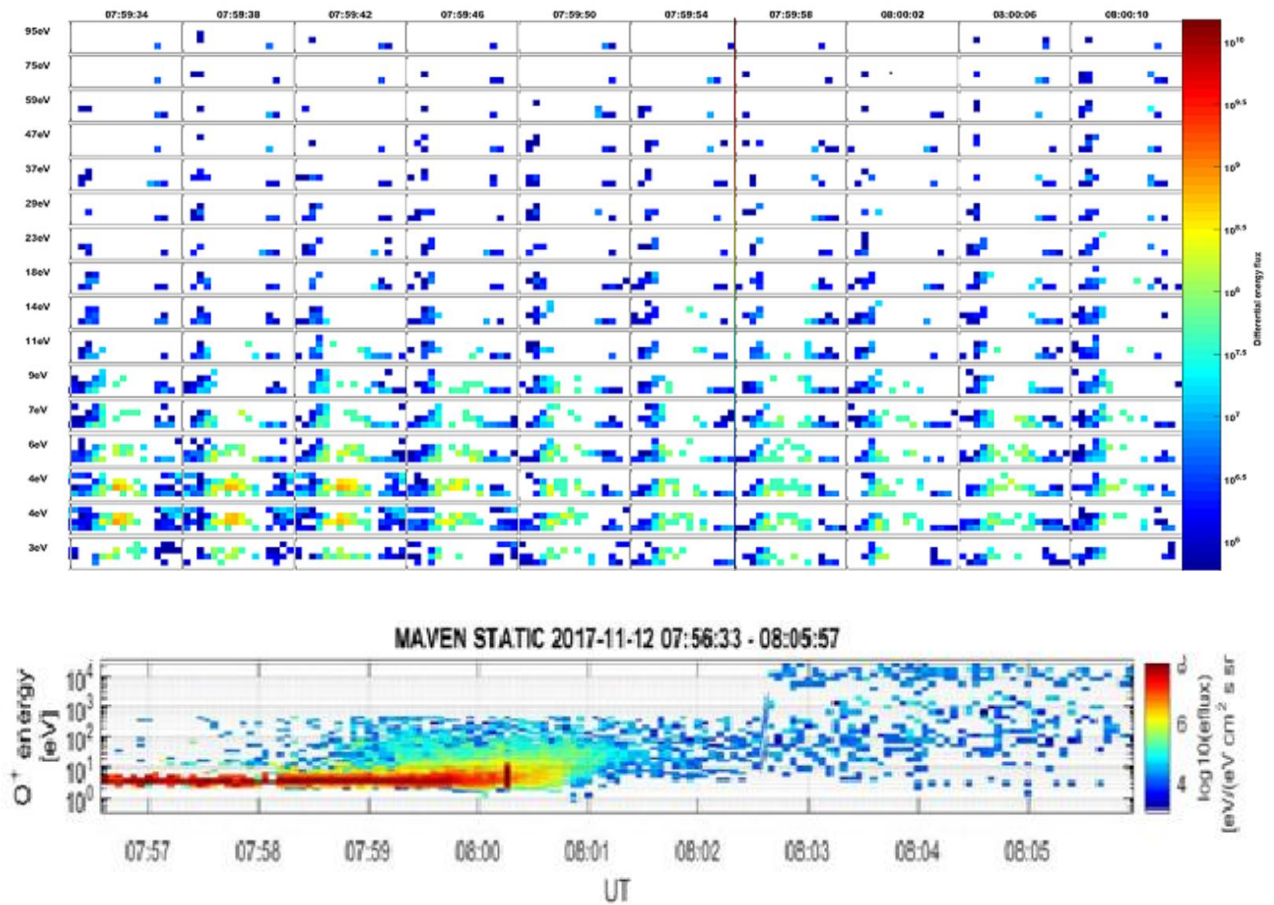

Figure 5. The O$^+$ ions velocity space distributions within time interval 07:59.34-08:00.10 UT on 12:11:2017 in the energy range from 3.0 eV at the bottom to 95 eV at the top. Each rectangle

covers 360⁰ azimuth span and -45⁰ -+45⁰ polar span. The cut of the energy-time spectrogram from Figure 4 (bottom) shows of what part of the pass the velocity distributions are shown.

Figure 5 shows velocity space distributions of $O^+$ ions observed within 40-seconds time interval corresponding to the pass of the heated ionospheric plasma. The center of each rectangle corresponds to the ram direction of the spacecraft. Two ion components are clearly seen: One component, located initially near the center of the rectangle in the energy range 3-11 eV is the ionospheric component. Two other components, located approximately at 60⁰ and 310⁰ azimuths, cover the energy range from 3 eV to 90 eV. The angular distance between these $O^+$ components diminishes at smaller energies suggesting that they converge at approximately zero velocity indicating that they are parts of one distribution. Another indication of this fact is the diminishing of their angular distances of at low energies. Indeed, if we have the ring or part of ring with the same width along it, the energetic part of the ring will have smaller angular span than less synergetic part.

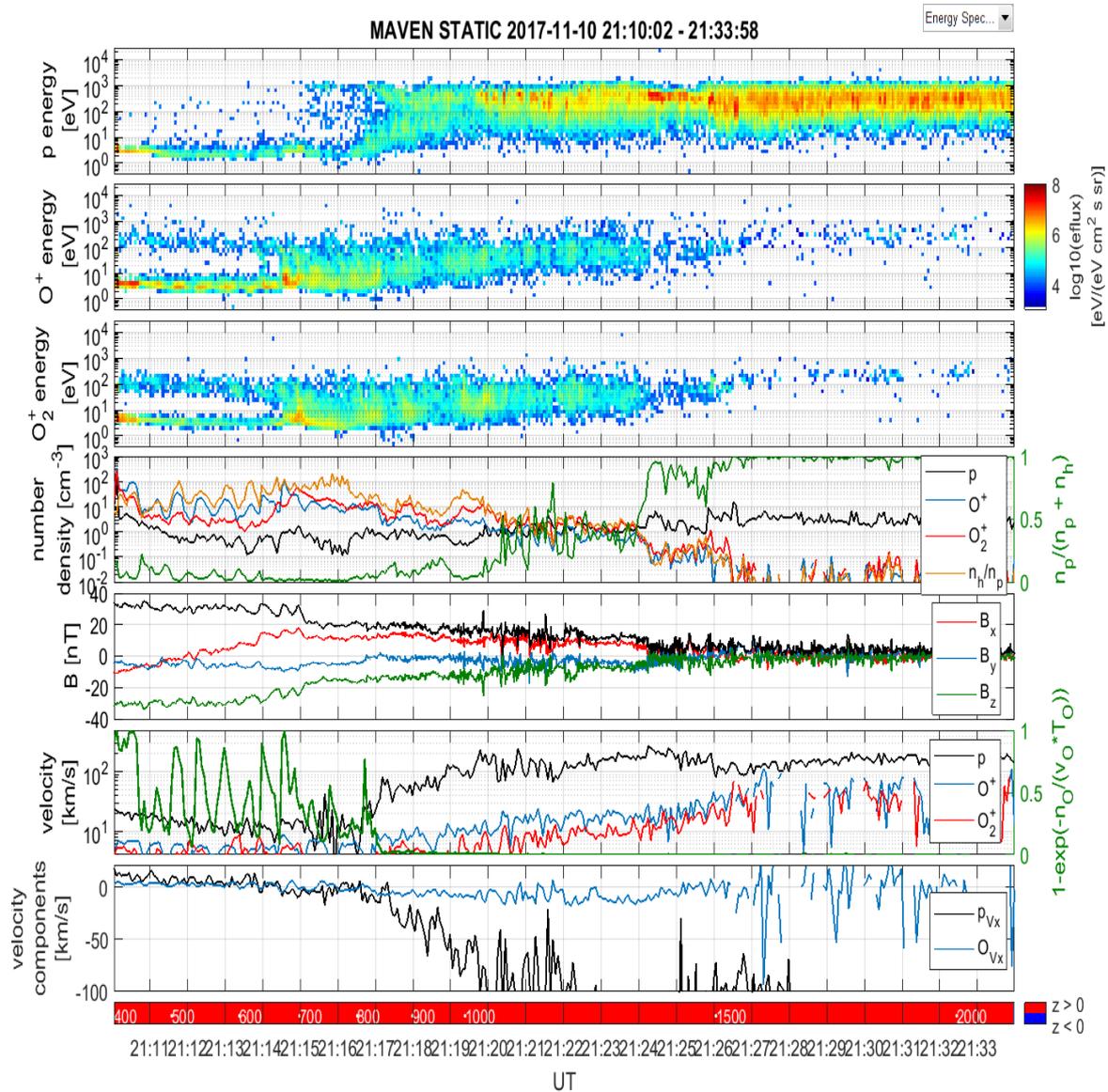

Figure 6. Another example of the interaction region crossing in case of the high energy of pick-up ions.

One simpler evidence of the heavy ions entrance in the obstacle to the solar wind is shown in Figure 6. The SZA of the MAVEN when it crossed the magnetopause was 35°. The pick-up of $O^+$ and $O_2^+$ ions during the flight of spacecraft through solar-wind-Mars interaction region at this time was very effective and pick-up ions were significantly more energetic.

Finally, Figure 7 is for March 13, 2016 at 13:42-13:57 UT pass through the magnetosheath, magnetosphere and ionosphere. This case shows clearly that ionospheric ions were strongly accelerated and heated in the region of energetic pick-up ions.

### Discussion and conclusion

Analysis of MAVEN observation of dayside interaction region of the solar wind with exosphere and ionosphere at the Solar-Zenith Angles ≤ 40° show that intermediate region forms between magnetosheath flow and ionosphere of Mars. This region does not have properties that are common for the boundary layer: it does not show gradual or turbulent transition. It usually has sharp external boundary with change of composition and often sharp change of velocity and other flow parameters. Mars has specific characteristics that combine fluid and kinetic properties that lead to formation of very specific layer not only associated with local fluid properties.

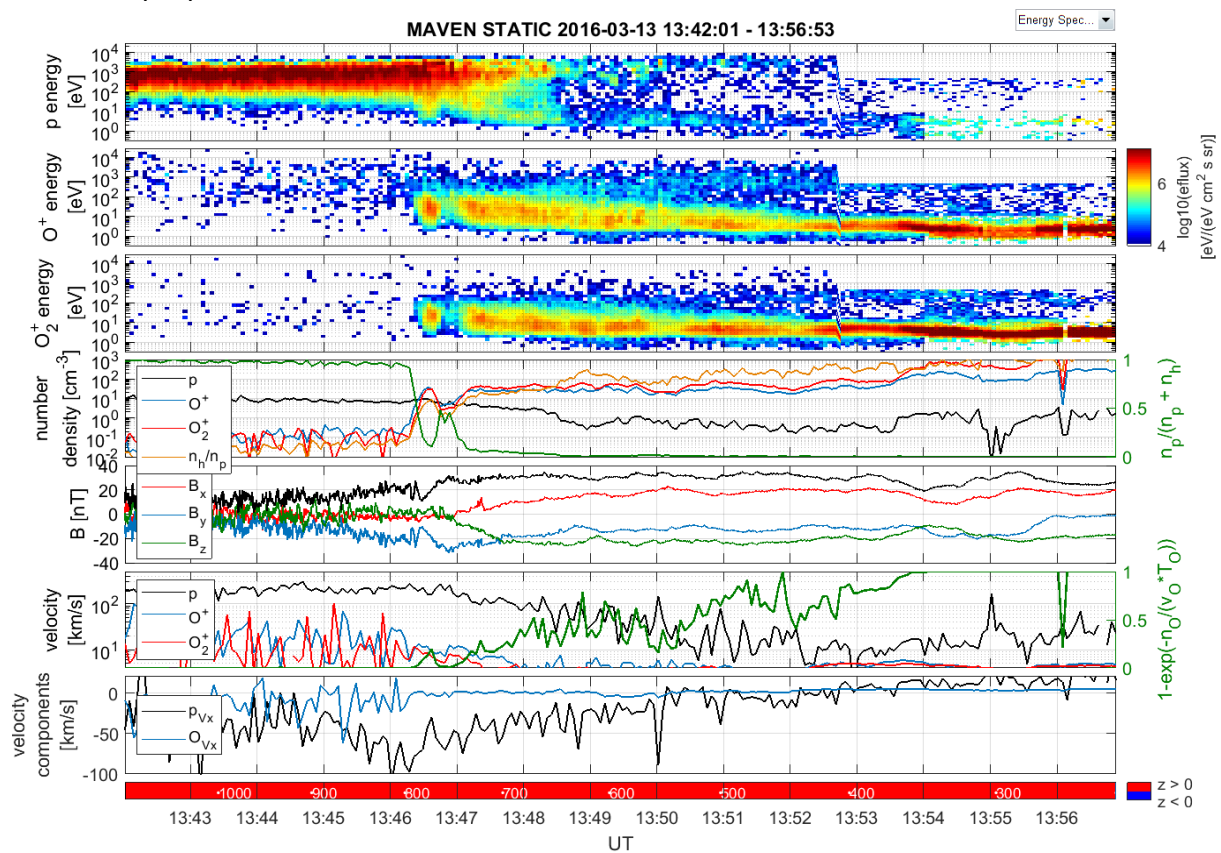

Figure 6. The pass of MAVEN on March 13, 2016 at ~ 13:42-13:57 UT at the SZA ~ 40°.

As we see from shown examples, the long range action is very important in formation of the region between the fluid-type external flow (magnetosheath) and ionosphere. The pick-up ions formed in the interaction between the solar plasma flow and photo-ionized heavy ions by means of electric field bring the energy to the upper atmosphere and ionosphere. The energy of these photo ions is spent in the same layer where the ionospheric electrons are heated and accelerated. The heating of the upper atmosphere needs to be involved in this process.

In this short communication we are limited to consideration of the structure and some phenomena apparently important for formation of the region between the shocked solar wind plasma around the Mars and Martian ionosphere. The name of the Magnetosphere may be appropriate for this region as it is topographically connected to the external part of Martian tail which is called magnetosphere.

More detailed analysis will be given elsewhere.